\documentclass[12pt]{iopart}
\usepackage{graphicx,iopams}
\usepackage{pstricks-add}
\usepackage{cite}
\usepackage[none]{hyphenat}

\usepackage{soul}

\def\pmb#1{\setbox0=\hbox{#1}
\kern-.025em\copy0\kern-\wd0 \kern.05em\copy0\kern-\wd0
\kern-.025em\raise.0433em\box0}

\newcommand{\text}[1]{\rm #1}

\begin{document}

\title[Marginal dimensions of the Potts model with invisible states]{Marginal dimensions
of the Potts model with invisible states}

\author{M. Krasnytska$^{1,2,4}$, P. Sarkanych$^{1,3,4}$, B. Berche$^{2,4}$, Yu. Holovatch$^{1,4}$, R. Kenna$^{3,4}$}

\address{$^{1}$
Institute for Condensed Matter Physics, National Acad.
Sci. of Ukraine, UA--79011 Lviv, Ukraine}

\address{$^{2}$
Institut Jean Lamour, CNRS/UMR 7198, Groupe de Physique Statistique,
Universit\'e de Lorraine, BP 70239, F-54506 Vand\oe uvre-les-Nancy
Cedex, France}

\address{$^{3}$
Applied Mathematics Research Centre, Coventry University, Coventry
CV1 5FB, United Kingdom}

\address{$^{4}$
Doctoral College for the Statistical Physics of Complex Systems,
Leipzig-Lorraine-Lviv-Coventry $({\mathbb L}^4)$}


\begin{abstract}

We reconsider the mean-field Potts model with $q$ interacting and
$r$ non-interacting (invisible) states. The model was recently
introduced to explain discrepancies between theoretical predictions
and experimental observations of phase transitions in some systems
where the $Z_q$-symmetry is spontaneously broken. We analyse the
marginal dimensions of the model, i.e., the value of $r$ at which
the order of the phase transition changes. In the $q=2$ case, we
determine that value to be  $r_c = 3.65(5)$; there is a second-order
phase transition there when $r<r_c$ and a first-order one at
$r>r_c$. We also analyse the region $1 \leq q<2$ and show that the
change from second to first order there is manifest through a new
mechanism involving {\emph{two}} marginal values of $r$. The $q=1$
limit gives bond percolation and some intermediary values also have
known physical realisations. Above the lower value $r_{c1}$, the
order parameters exhibit discontinuities at temperature $\tilde{t}$
below a critical value $t_c$. But, provided $r>r_{c1}$ is small
enough, this discontinuity does not appear at the phase transition,
which is continuous and  takes place at $t_c$. The larger value
$r_{c2}$ marks the point at which the phase transition at $t_c$
changes from second to first order. Thus, for $r_{c1}< r < r_{c2}$,
the transition at $t_c$ remains second order while the order
parameter has a discontinuity at $\tilde{t}$. As $r$ increases
further, $\tilde{t}$ increases, bringing the discontinuity closer to
$t_c$. Finally, when $r$ exceeds $r_{c2}$ $\tilde{t}$ coincides with
$t_c$ and the phase transition becomes first order. This new
mechanism indicates how the discontinuity characteristic of first
order phase transitions emerges.
\end{abstract}

 \submitto{\JPA}

\eads{\mailto{kras$_-$mariana@icmp.lviv.ua},
\mailto{petrosark@icmp.lviv.ua}
\mailto{bertrand.berche@univ-lorraine.fr},
\mailto{hol@icmp.lviv.ua}, \mailto{R.Kenna@coventry.ac.uk}}

\maketitle


\vskip 1cm

\section{Introduction}\label{I}

The concept of universality - wherein certain features of a system
do not depend on its details - plays a central role in understanding
physical properties of various multi-particle systems. Continuous
(second-order) phase transitions provide examples of phenomenon
which exhibit universality \cite{Stanley88,Domb96}. For short-range
interacting systems, global factors, such as the dimension of space
or the dimensions and symmetries of  order parameters, define
universal features of such a phase transition. These features are
shared by different systems, independently of the details of the
system structure. Such systems then belong to the same {\em
universality class}. This is usually identified by critical
exponents, critical amplitude ratios and scaling functions. Another
universal quantity, which is intrinsic to the critical behaviour of
complex systems, is the {\em marginal dimension} which marks the
number of order-parameter components for which the order of the
phase transition changes.

Examples of such systems include  the $O(m)$-symmetric spin models
\cite{Stanley69}. The transitions there are of second order provided
that the space dimension exceeds a lower-critical value $d>d_{lc}$
(where $d_{lc}=1$ for the Ising case $m=1$ and $d_{lc}=2$ for $m\geq
2$) \cite{Mermin66,Hohenberg67}. However, when the $O(m)$ symmetry
is broken by the presence of terms invariant under the cubic group
(the so-called anisotropic cubic model, relevant for an account of
crystalline anisotropy \cite{Aharony76,cubic}), it leads to the
emergence of a marginal dimension $m_c$. For given space dimension
$d$, $m_c(d)$ separates regions where the phase transitions of a
given symmetry are of first or of  second order. For example, the
$d=3$ cubic crystal with three easy axes should undergo either a
first- or second-order phase transition provided $m_c$ is less than
or greater than 3, respectively. Theoretical estimates are
in favour of $m_c(d=3)<3$ \cite{Folk00,Dudka04}, supporting the
first-order scenario in these systems \cite{cubic}. Another  example
is the $q$-state Potts model \cite{Potts52}. Since this model has a
discrete symmetry group, $Z_q$, the lower critical dimension is
{$d_{lc}=1$}. The marginal value $q_c(d)$, for $d>d_{lc}$, separates
the first- and second-order regimes there too. For $d=2$ the
transition is of second order for $q\leq q_c = 4$  and of first
order otherwise. For $d=3$, the marginal value  $q_c$ is below $3$
\cite{Wu82}. More examples of marginal dimensions that separate
regions of phase transitions of different types are given by systems
with non-collinear ordering \cite{non-colinear}, frustrations
\cite{frustrations}, structural disorder
\cite{structural_disorder,Dudka04}, and competing fluctuating fields
\cite{competing_fields}.

In this paper we will be interested in marginal dimensions of the
Potts model with invisible states \cite{Tamura10}. The model has
been recently introduced in order to explain discrepancies between
theoretical predictions and experimental observations of the phase
transition in some two-dimensional systems where the $Z_3$-symmetry
is spontaneously broken \cite{example}. Although, these systems
undergo a ferromagnetic phase transition, it appears to be of first
order, whereas a standard Potts model predicts the second order
scenario at $d=2$, $q=3$. The model continues to attract
considerable interest
\cite{Tanaka11a,Tanaka11b,Mori12,Tamura12,Ananikian13,Enter11a,Enter11b,Johnston13},
although, as we will show below, some principal questions about its
behaviour remain unsolved. In particular, to explain changes in the
marginal number of states (marginal dimension) $q_c$ that separates
the first and second order regimes, the model introduces $r$
additional Potts states (so-called invisible states)  that  do not
contribute to system's interaction energy but do contribute to the
entropy. Hence, for  fixed space dimension $d$ and number of states
$q$, the value of $r_c$ represents a border separating the first-
and second-order regimes.

The question of the marginal dimensionality of the Potts model with
invisible states is one of the central issues discussed within the
context of this model. The mean-field analysis in the framework of
the Bragg-Williams approximation of Refs.
\cite{Tamura10,Tanaka11b,Tamura12} lead to an estimate $3<r_c<4$ for
$q=2$. Obtained, as it is, within the mean-field approach, this
estimate does not carry a dependence on the space dimensionality. A
different mean-field approach employing 3-regular random (thin)
graphs also demonstrates a change in the order of the phase
transition, but the value of $r_c\simeq17$ obtained at $q=2$
\cite{Johnston13} is much higher than in the Bragg-Williams case.
Numerical simulations of the Potts model with invisible states in
$d=2$ dimensions gave solid evidence that the model exhibits a
first-order phase transition at $q=2,3,4$ for high values of $r$ but
it still remains a challenge for numerics to get a more precise
estimate of $r_c$ \cite{Tamura10}. Rigorous results prove the
existence of a first-order regime for any $q>0$, provided that $r$
is large enough \cite{Enter11a,Enter11b}. Exact results for the
value of $r_c$ for the model are known for a Bethe lattice
\cite{Ananikian13} too.

In this paper we  use the mean-field approach to determine more precise estimates for $r_c$ when $q=2$ and  to find $r_c(q)$ in the region $1\leq q < 2$.
The aim is to  deliver a better understanding of  how $r_c$ changes with $q$.
The region also includes the physically accessible case of bond percolation $q=1$ which is of special interest in its own right.

The rest of this paper is organized as follows.
In  Section~\ref{II} we obtain the expression for the free energy of the Potts model with  invisible states in the mean-field approximation.
The analysis of marginal dimensions is presented in Section~\ref{III}.
In the concluding section, we summarize the results obtained.

\section{Free energy in the mean-field approximation}\label{II}

The Hamiltonian of the $(q+r)$-state Potts model with $r$ invisible
states reads \cite{Tamura10}
\begin{equation}\label{1}
{\cal H}=-\sum_{<i,j>}\delta _{S_i,S_j}\sum_{\alpha=1}^q \delta
_{S_i,\alpha}\delta _{S_j,\alpha}- h\, \sum _i \delta_{S_i,1},
\end{equation}
where, $S_i=1,\dots,q, (q+1),\dots,(q+r)$ is a Potts spin variable
on a site $i=1,\dots,N$, $\delta_{a,b}$ are Kronecker deltas { and
$h$ is a magnetic field acting on the first visible state}. The
first sum in the first term spans all distinct nearest neighbour
pairs. Only states with $S_i=1,\dots,q$ contribute to the
interaction term in the Hamiltonian and without loss of generality
we may put the coupling constant there equal to 1. The remaining $r$
states do not contribute to the interaction energy
 but they increase the number of configurations available, and hence they contribute to the entropy
(as well as the internal energy). An external field $h$ is chosen to
favour the $S_i=1$ state.

As mentioned above,  a Bragg-Williams type mean-field calculation and Monte-Carlo simulations in two dimensions lead to the conclusion that with increasing $r$ the phase transition in the model (\ref{1}) becomes  `harder';  the second-order transition changes to a first-order one.
As $r$ is further increased, the latent heat and the jump in the order parameter also increase at the first-order phase-transition point \cite{Tamura10,Tanaka11b,Tamura12}.
Since in the mean-field approximation the transition of the usual $q$-state Potts model is of  first order for $q>2$, mean-field analysis of Refs.~\cite{Tamura10,Tanaka11b,Tamura12} concentrated on the Ising case
$q=2$ to demonstrate the change in the order of the phase transition with increasing of $r$.
It was shown that the transition becomes  first order for $r>3$.
However the value of $r_c$ has not yet been determined precisely for $q=2$.
Here, we will apply another variant of the mean-field approach to obtain precise estimates for the value of the marginal dimension $r_c$ for $q=2$ as well as to analyse the entire $1\leq q \leq 2$ region.
Besides the Ising case, this region includes another frequently investigated physically relevant case, namely that of  bond percolation, $q=1$ \cite{Wu82}.
Another challenge is to observe the continuous change of the marginal dimension $r_c(q)$, both because analytic continuation in the number of states is an inherent feature of the field-theoretical description of critical phenomena and because the Potts model at non-integer $q$ is relevant for the description of a number of interesting physical phenomena.
To give just a few examples, besides the bond percolation previously mentioned, the limit $q\rightarrow 1$ describes turbulence \cite{Fortiun69,Giri77}.
Universal spanning trees (Fortuin-Kasteleyn graphs) are described by a zero-state $q=0$ Potts model \cite{Stephen76,Deng}.
This limit is related to an arboreal gas model \cite{Jacobsen05} and resistor networks \cite{Fortiun72,Stenull}.
The Potts model at $q=1/2$ corresponds to a spin glass model \cite{Aharony78,Aharony79}, which can be also
used to analyze the evolution of syntax and language \cite{Siva}.
Finally, the Potts model in the region $0\leq q<1$ describes gelation and vulcanization processes in branched polymers \cite{Lubensky78}.

To proceed with the mean-field analysis, let us introduce thermodynamic averages
  \begin{eqnarray}\label{2}
 \langle \delta
_{S_i,\alpha} \rangle=\left\{
\begin{array}{ccl}
                & \mu \, ,   & \alpha=1, \\
                & \nu_1 \, , & \alpha=2, \dots, q, \\
                & \nu_2\, ,  & \alpha=q+1, \dots, r\, .
              \end{array}
  \right.
\end{eqnarray}
Here, the averaging is performed with respect to the Hamiltonian (\ref{1})
\begin{equation}\label{3a}
\langle\dots\rangle = \frac{1}{\cal Z}{\rm Tr}\, (\dots)
e^{-\beta{\cal H}}, \quad \quad {\mbox{with}} \quad \quad
 {\cal Z} =  {\rm Tr}\, e^{-\beta{\cal H}},
\end{equation}
in the thermodynamic limit, where $\beta$ is the  inverse temperature and the trace is taken over all
possible spin configurations.

By (\ref{2}), in the spirit of the mean-field approximation, we assume that the mean value of a spin in a given state does not depend on its coordinate.
Note, that three different averages $\mu$, $\nu_1$, and $\nu_2$ are necessary to take into account the state
favoured by the magnetic field and to discriminate between visible and invisible states.
Their high- and low- temperature asymptotics are given in Table~\ref{tab1}.
At high temperatures all states are equally probable, whereas at low temperatures the direction of
symmetry breaking is determined by the direction of the magnetic field.

\begin{table}[b]
\caption{Low and high temperature asymptotics of the thermodynamic
averages, Eq.~(\ref{2}), and for the order parameters, Eq. (6).
\label{tab1}}
\begin{center}
    \begin{tabular}{|c|c|c|c|c|c|}
      \hline
       $\beta \rightarrow \infty$ & $\mu=1$ & $\nu_1=0$ & $\nu_2=0$ & $m_1=1$ & $m_2=1$ \\  \hline
      \hline
      $\beta\rightarrow 0$ & $\mu=\frac{1}{q+r}$ & $\nu_1=\frac{1}{q+r}$ & $\nu_2=\frac{1}{q+r}$ &
      $m_1=0$ & $m_2=0$ \\
      \hline
    \end{tabular}
\end{center}
\end{table}

This asymptotic behaviour together with an obvious normalization condition:
\begin{equation}\label{4}
\mu+(q-1)\nu_1+r\nu_2=1
\end{equation}
allows one to define the order parameters:
\begin{eqnarray}\nonumber
 m_1&=&\mu-\nu_1, \\ \label{5}
 m_2&=&\mu-\nu_2.
  \end{eqnarray}
Both $m_1$ and $m_2$  exhibit standard temperature asymtotics in that they vanish for $\beta \rightarrow 0$ and are equal to one for $\beta \rightarrow \infty$, see
Table~\ref{tab1}.
However, as we will see below, only $m_1$ has a physical interpretation as a quantity that appears below the transition point and breaks the system symmetry.
It is easy to show that also the following conditions are satisfied:{
\begin{eqnarray}\nonumber
 \mu&=&\frac{m_2r+m_1q+1-m_1}{q+r},\\ \nonumber
 \nu_1&=&\frac{(m_2-m_1)r+1-m_1}{q+r},\\ \label{6}
 \nu_2&=&\frac{(m_1-m_2)q+1-m_1}{q+r}.
  \end{eqnarray}}

The first  Kronecker $\delta$ in the Hamiltonian (\ref{1}) is rendered redundant by the other two, so that
 \begin{equation}\label{8}
{\cal H} =- \sum_{<i,j>}\sum_{\alpha=1}^q \delta _{S_i,\alpha}\delta
_{\alpha,S_j}-h \sum _i  \delta_{S_i,1}.
  \end{equation}
To obtain the mean-field Hamiltonian, we represent each Kronecker-delta term as a sum of its mean value and deviation from that mean.
Neglecting terms comprising  a product of two such deviations,
 \begin{equation}\label{9}
{{\cal H}=-\frac z2\sum_{i}[\mu(2\delta
_{1,S_i}-\mu)+\sum_{\alpha=2}^q (2\delta
_{\alpha,S_i}-\nu_{1})\nu_{1}]-h\sum_i \delta_{S_i,1},}
  \end{equation}
  where $z$ is the number of the nearest neighbours.
For the partition function (\ref{3a}) we then get
\begin{equation}\label{10}
\nonumber {\cal Z}\,=\,
e^{-N\beta {z}(\mu^2+(q-1)\nu_{1}^2)/2}
\prod_i[e^{\beta(h+z\mu)}+(q-1)e^{\beta z\nu_{1}}+r].
  \end{equation}
We consequently derive the free energy per site as
\begin{eqnarray}
\nonumber
 f(m_1,m_2)&=& -\frac{1}{\beta N}\log {\cal Z} =
\frac{z}{2} \Bigg(\frac{(m_1q-m_1+m_2r+1)^2}{(q+r)^2}+ \\
\nonumber &&
\frac{(q-1) (-m_1r-m_1+m_2r+1)^2}{(q+r)^2}\Bigg)-\\
\nonumber  && \frac{1}{\beta}\log \Bigg\{(\exp\Bigg[  {\beta
\left(h+\frac{ z (m_1q-m_1+m_2r+1)}{q+r}\right)}\Bigg] +\\ &&
\label{11}
 (q-1) \exp\Bigg[  {\frac{ \beta  z
(-m_1r-m_1+m_2r+1)}{q+r}}\Bigg] +r\Bigg\}.
\end{eqnarray}
For $r=0$ one recovers the free energy of the standard Potts model
as a function of a single order parameter $m_1$ in the mean-field
approximation \cite{mfa}. Of course,  $m_2$ does not arise in the
standard Potts model. There the transition is of the second order
only if $q\leq 2$. In the following, we are interested how the
presence of invisible states changes the order of this transition.

\section{Phase transition and marginal dimensions}\label{III}

\begin{figure}[t]
\centerline{\includegraphics[angle=0, width=8cm]{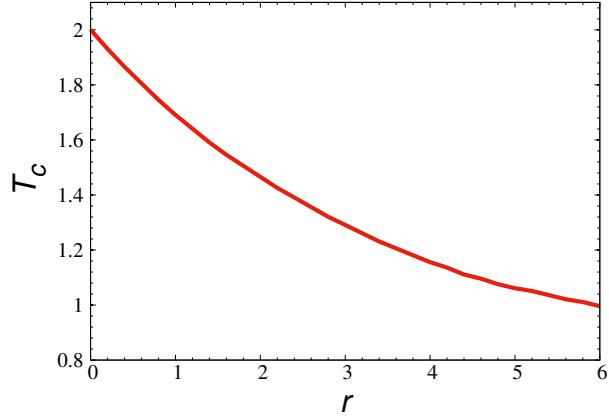}}
 \caption{Critical temperature of the Potts model for $q=2$, $z=4$ as a function of the number of invisible states $r$.
The critical temperature is a smooth function of $r$ and tends to
zero as $r\to\infty$. \label{fig1}}
\end{figure}

With the expression (\ref{11}) to hand, the thermodynamics of the model are obtained via minimization of the free energy with respect to the two parameters $m_1$ and $m_2$.
In particular, the system of equations that determines the  free energy extrema, ${\displaystyle{\partial f/\partial m_1 =
\partial f/\partial m_2=0}}$, reads
\begin{eqnarray}
\!\!\!\!\!\!\!\!\!\!\!\!\!\!
\frac{(q+r) \left[{e^{\beta  (h+ m_1 z)}-r-1}\right]}{e^{\beta  (h+ m_1 z)}+r e^{\frac{ \beta  z (m_1
r+m_1-m_2 r-1)}{q+r}}+q-1}
& = &
m_1[q+r (r+2)]-r (m_2 r+1) \, , \quad \quad  \label{12a} \\
\!\!\!\!\!\!\!\!\!\!\!\!\!\!  \frac{(q+r) \left[{e^{\beta  (h+ m_1
z)}+q-1}\right]}{e^{\beta  (h+ m_1 z)}+r e^{\frac{ \beta  z (m_1
r+m_1-m_2 r-1)}{q+r}}+q-1} & = & -m_1 r(q-1)+m_2 q r+q\, . \label{12b}
\end{eqnarray}
The solutions of these equations, $m_1(T,h)$, $m_2(T,h)$ are further
analysed to ensure they meet the condition of stability, i.e that
they correspond to the free energy minimum, or to local minima in
the case of a first-order transition. From these considerations,
and numerically solving the system of non-linear equations (\ref{12a}), (\ref{12b}), we find two types of solutions at zero
external magnetic field and finite temperature, namely (i)
$m_1(T,0)=0,\, m_2(T,0)\neq 0$ and (ii) $m_1(T,0)\neq 0,\,
m_2(T,0)\neq 0$. Note that $m_2(T,0)$ never vanishes at finite
temperature. Therefore only $m_1(T,0)$  is a proper order parameter,
delivering a spontaneous magnetization that signals the occurrence
of a phase transition. For fixed $q$, the transition from solution
(i) to (ii) occurs at a finite $r$-dependent temperature $T_c(r)$.

\subsection{The case $q=2$}
First let us consider the extension of the Ising model with
invisible states. In Fig.~\ref{fig1}  we plot the transition
temperature for $q=2$ as a function of $r$ for $z=4$. The critical
temperature is a smooth function of $r$ and tends to zero as
$r\to\infty$. In this limit the system becomes one of
non-interacting particles. Note that for the $2D$ Potts model with
invisible states on a square lattice  $T_c$ vanishes for large
$(q+r)$ as $T_c\approx 2/\ln(q+r)$\cite{Enter11a}.

\begin{figure}[t]
\centerline{\includegraphics[angle=0,
width=7.5cm]{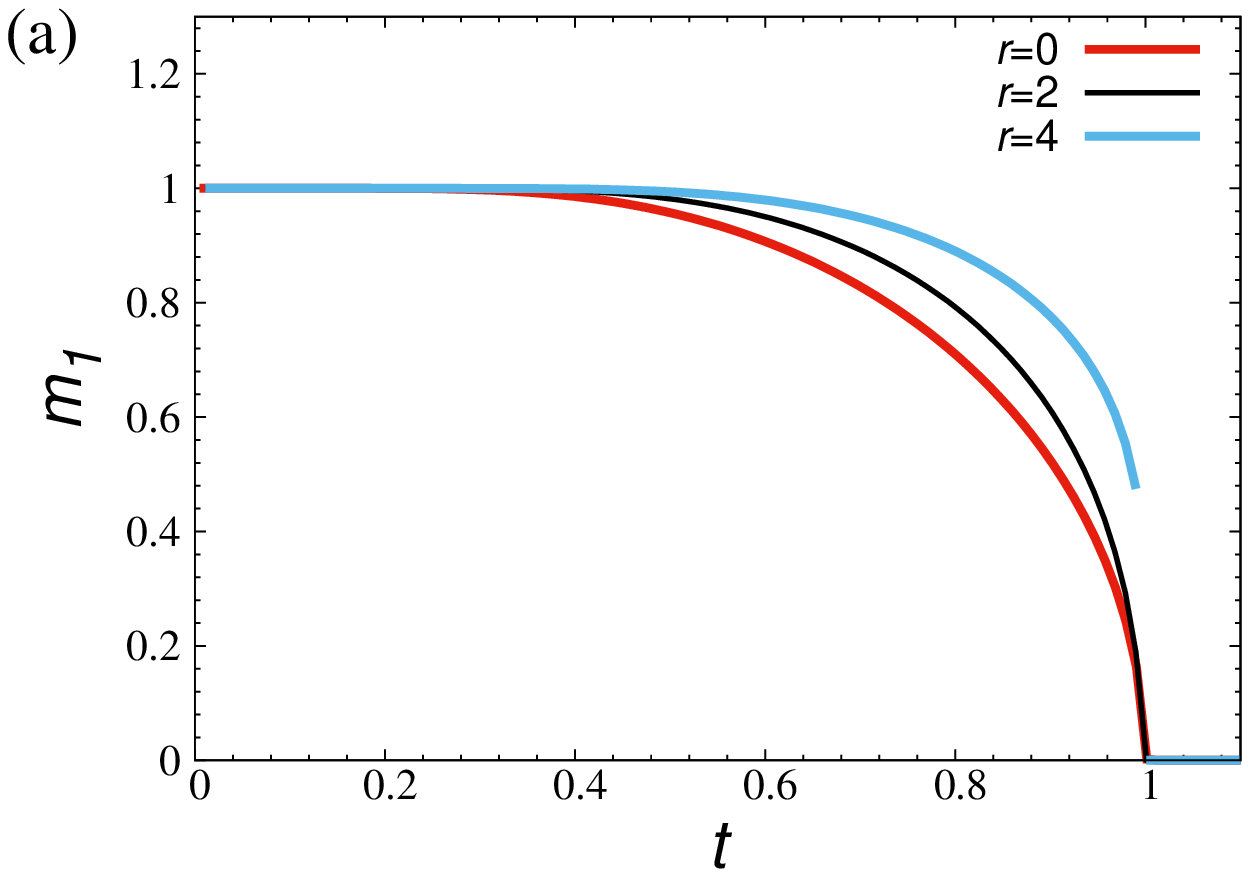}\hspace{2em}
\includegraphics[angle=0, width=7.5cm]{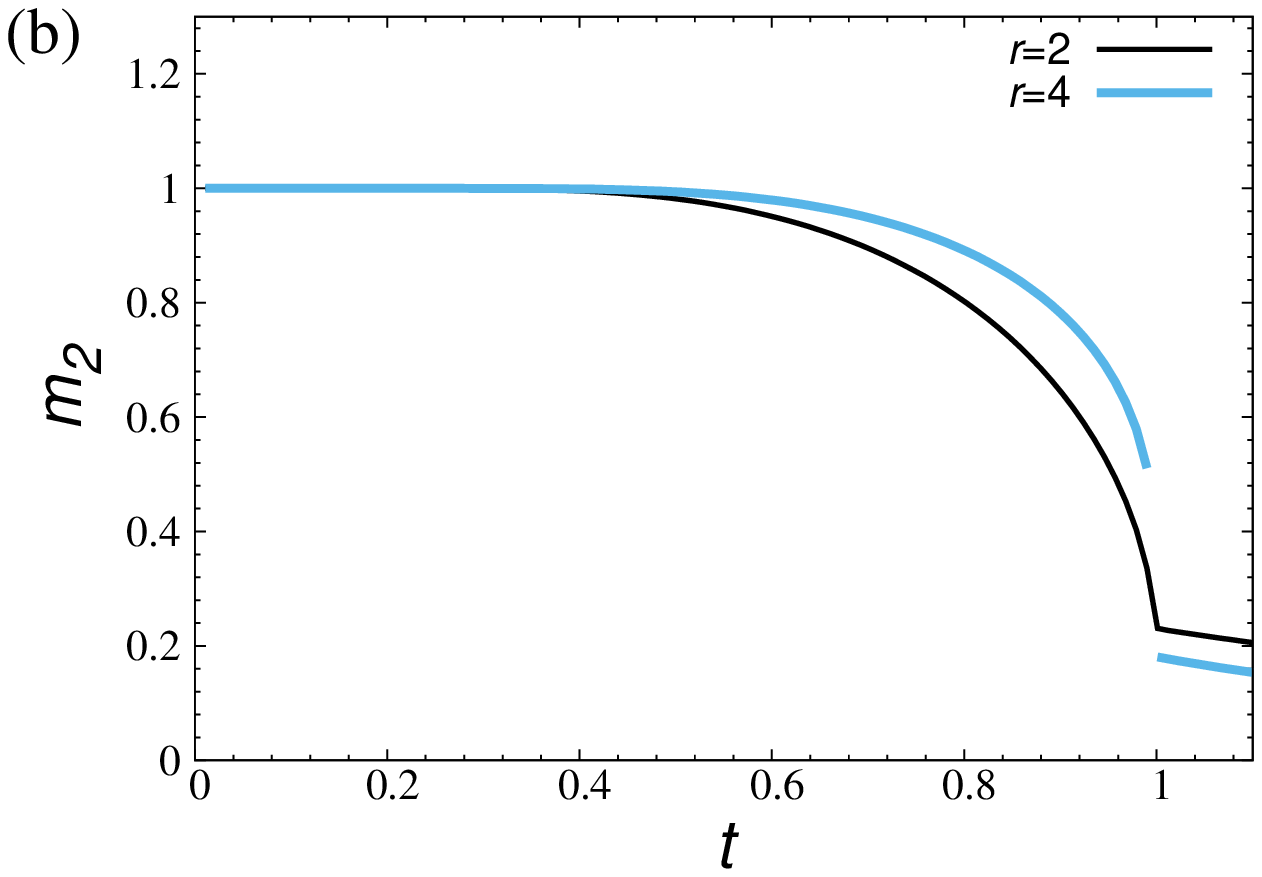}}
 \vspace{1ex} \caption{The  dependency of the order parameters $m_1$ [Panel (a)]
and $m_2$ [Panel (b)]  on the reduced temperature $t=T/T_c$ for
$r=0, 2, 4.$ For $r=0$, we only have a single order parameter,
namely $m_1$. Indeed, $m_1(t)$ is a proper order parameter in that
it vanishes on one side of the phase transition. In contrast,
$m_2(t)$ never vanishes at finite temperature. However both $m_1$
and $m_2$ can be used to distinguish between the first and the
second-order regimes as the plots demonstrate.
 \label{fig2}}
\end{figure}

As has been shown in \cite{Tamura10}, $r=4$ invisible states are
sufficient to change the phase transition of the $q=2$ Potts model
from  second  to  first order. This sets an upper bound for the
marginal dimension as $r_c(q=2)<4$. We display the temperature
dependence of the  order parameters in Fig.~\ref{fig2}  for
$r=0,2,4$. Since $T_c$ is $r-$dependent we use the reduced
temperature $t=T/T_c$. As we have noted before, one only has
$m_1(T)$ for $r=0$. Depending on the value of $r$, the temperature
dependency of both order parameters $m_1$, $m_2$ is characterized by
two different regimes. For $r=0, 2$ the plots are continuous,
signalling second-order phase transitions. However when $r=4$ we
observe a jump at the critical temperature. Note that both $m_1$ and
$m_2$ may be used to distinguish between the first and the second
order regimes. However, it is worth re-emphasising that above the
critical temperature $m_1(t)=0$, while $m_2(t)$ vanishes only for
the infinite temperature.

\begin{figure}[t]
 \centerline{\includegraphics[angle=0, width=8cm]{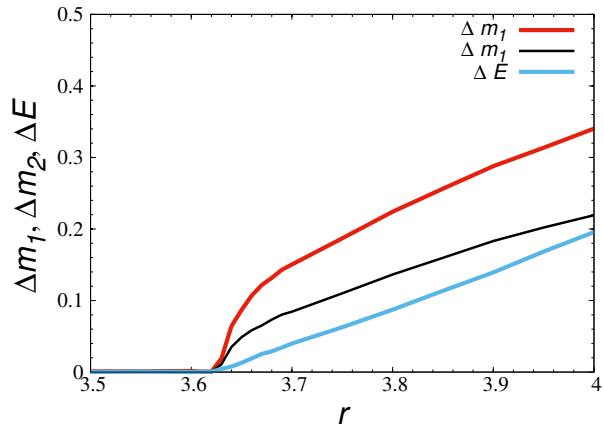}}
\caption{Jumps in the order parameters $\Delta m_1$ (red curve),
$\Delta m_2$ (black curve) and the latent heat $\Delta E$ (blue
curve) of the Potts model at $q=2$  as functions of the number of
invisible states $r$.
  \label{fig3}}
\end{figure}

To locate the marginal value $r_c$, we define the jump in the order parameters by
\begin{equation}
\label{cond}
\Delta m_j=\lim_{t \to 1^-}
m_j(t)-\lim_{t \to 1^+} m_j(t), \hspace{2em} j=1,2,
\end{equation}
and analyse the behaviour of $\Delta m_j$ as function of $r$. The
first appearance of a non-zero value of $\Delta m_j$ corresponds to
the onset of the first-order phase transition. In  Fig.~\ref{fig3}
we plot $\Delta m_1$ and $\Delta m_2$ as functions of the number of
invisible states $r$. Similar behaviour is observed for the latent
heat $\Delta E=-\Delta S\, T_c$, where $\Delta S$ is the entropy
jump at the phase transition point:
\begin{equation}
\label{cond2} \Delta S=\lim_{t \to 1^-} S(t)-\lim_{t \to 1^+}
S(t).
\end{equation}
This is also plotted in Fig.~\ref{fig3}. The values of $r_c$
obtained from the vanishing of $\Delta m_1$, $\Delta m_2$, and
$\Delta E$ are: $r_c=3.629(1)$, $r_c=3.627(2)$, and $r_c=3.617(3)$,
respectively. Averaging these values we get $r_c=3.622(8)$. This
estimate agrees well with the $z\to \infty$ limit of the result
obtained for the Potts model with $q=2$ visible and $r$ invisible
states on the Bethe lattice  with $z$ nearest neighbours
\cite{Ananikian13}: $r_c=
\lim_{z\to\infty}\frac{4z}{3(z-1)}\Big(\frac{z-1}{z-2}\Big)^2\simeq
3.62$.

In the vicinity of $r_c$, the order parameter jumps can be
approximated by a power-law decay:
\begin{equation}
\Delta m_1\sim (r-r_c)^{a_1},\hspace{1cm}\Delta m_2\sim
(r-r_c)^{a_2}. \label{scaling}
\end{equation}
Numerical fits in the interval $r=3.625$ to $4.0$ yield estimates
for the exponents: $a_1=0.477(10)$ and $a_2=0.566(15)$.

\subsection{The case $1\leq q<2$}
\begin{figure}[t]
\centerline{\includegraphics[angle=0, width=8cm]{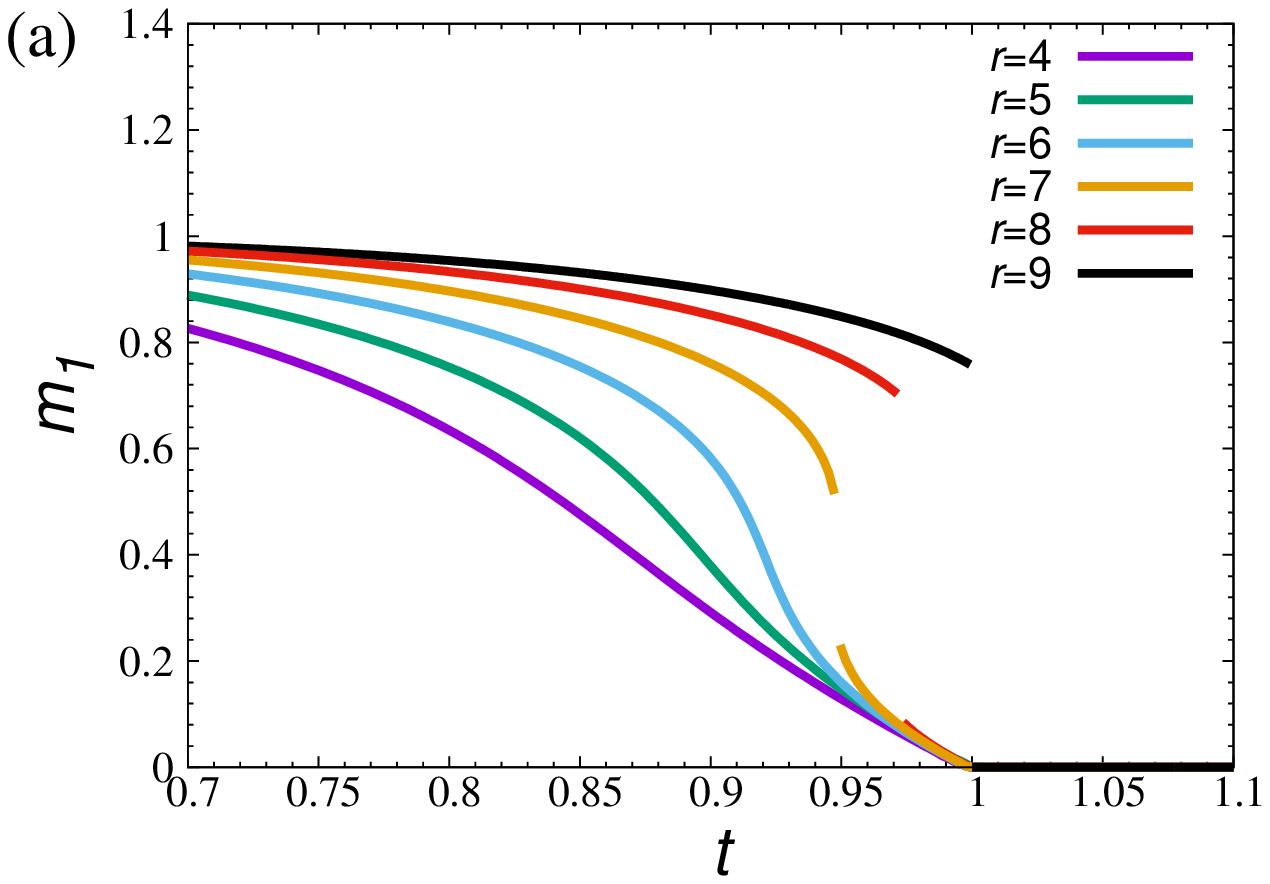}
\includegraphics[angle=0, width=8cm]{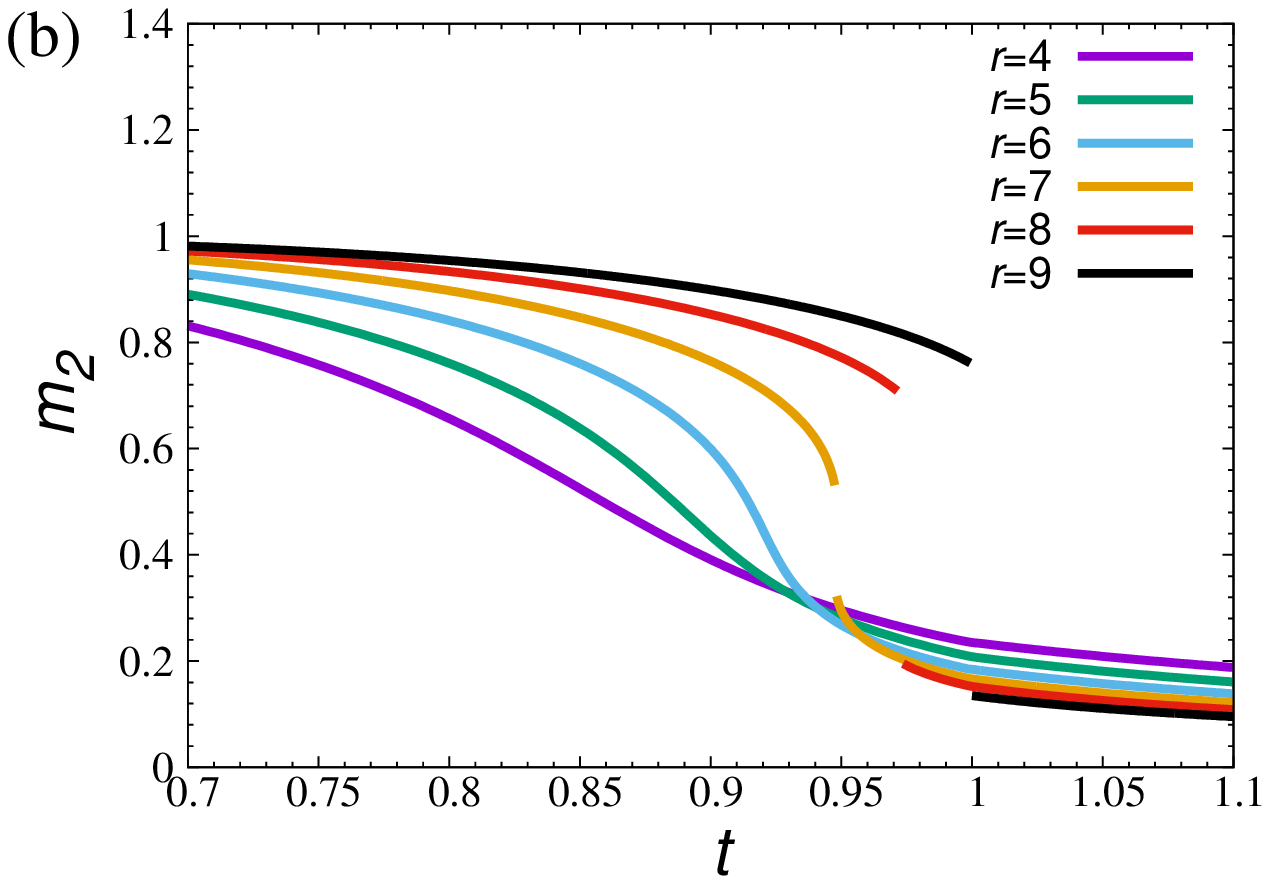}}
\vspace{2ex}
   \centerline{\includegraphics[angle=0, width=8cm]{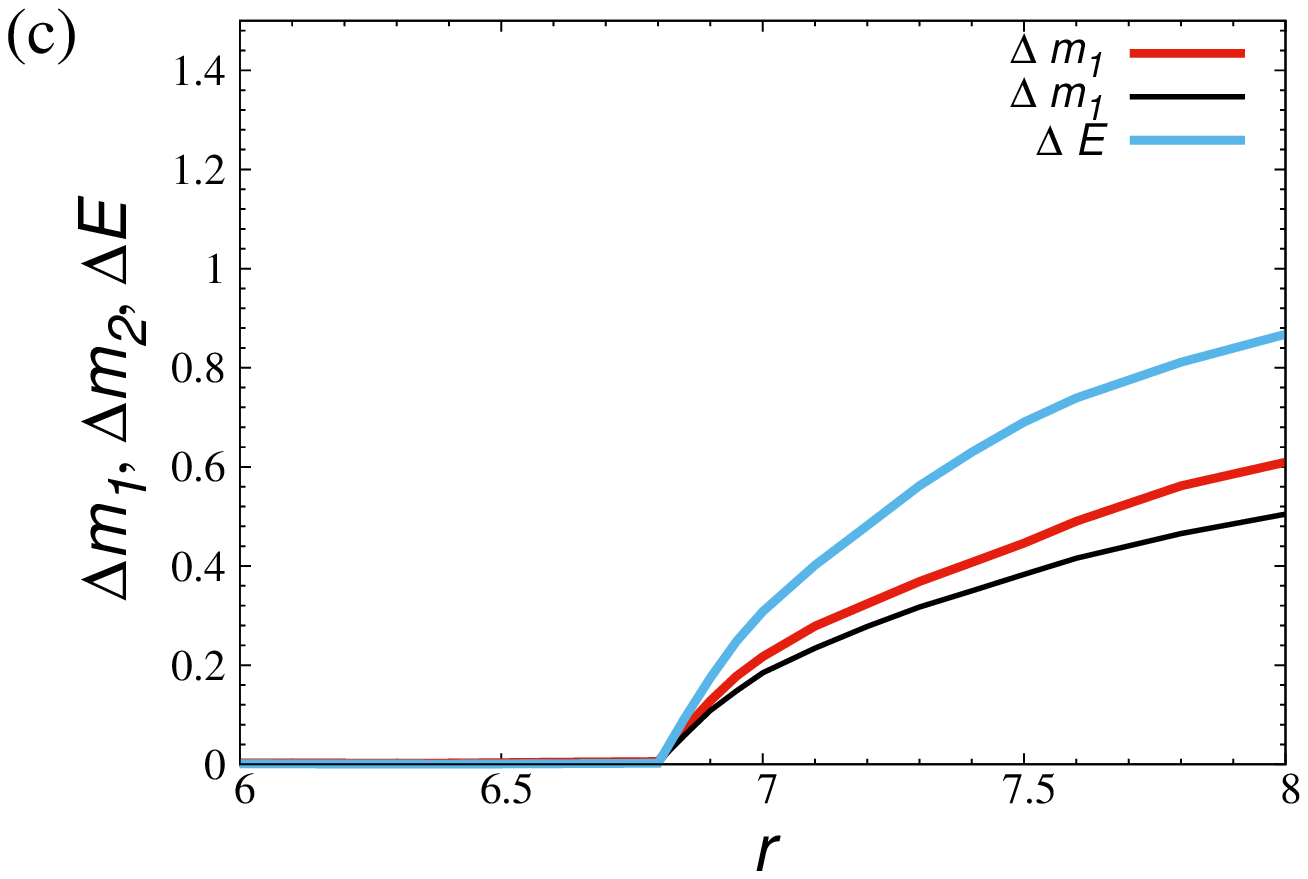}}
\vspace{3ex} \caption{Dependencies of the first (a) and second (b)
order parameters
 on the reduced temperature $t$ for $r=4,5,6,7,8,9$ at
$q=1.2$. The second-order phase transition transforms to a first
order phase transition at $r_{c_2}\simeq 8.495(5)$. Panel (c):
discontinuities in the order parameter  $\Delta m_1$ (red curve),
$\Delta m_2$ (black curve) and the latent heat $\Delta E$ (blue
curve) as functions of $r$ for $q=1.2$ at $t=\tilde{t}$.
\label{fig4}}
\end{figure}

Let us consider now the region $1\leq q<2$. Typical behaviour of the
order parameters $m_1(t)$ and $m_2(t)$ for  fixed values of $q$ is
shown in Fig.~\ref{fig4}. There we plot these functions  for $q=1.2$
and $r=4,5,6,7,8,9$. For small values of $r$, $m_1(t)$ and $m_2(t)$
are smooth functions of $t$ and the transition is second order. We
observe that $m_1(t)$   vanishes linearly as $t$ approaches $t_c=1$
from below. This corresponds to the familiar mean-field result for
the percolation critical exponent $\beta=1$.

For larger values of  $r$, and starting from a certain value
$r=r_{c1}$, gaps in $m_1(t)$ and $m_2(t)$ appear at $\tilde{t}<t_c$.
The marginal dimension $r_{c1}$ obtained from the vanishing of these
functions is $r_{c1}\simeq 6.834(11)$. The occurrence of this gap
does not affect the order of the phase transition occurring at
$t_c=1$, which, for these values of $r$, remains second order
because the order parameters remain continuous there. The gap at
$\tilde{t}$ increases with further increases of $r$ and finally, at
$r=r_{c_2}$, $\tilde{t}$ and $t_c$ coincide. It is at this point
that the transition at $t_c=1$ becomes first order. The value
$r_{c_2}$ is therefore the marginal dimension (i.e., it is the value
of $r$  at which the phase transition changes its order). It is
defined by the condition
\begin{equation}\label{cond1}
\Delta m_1>0 \hspace{1em}{\rm at}\hspace{1em} m_1(t\to t_c^+)=0\,.
\end{equation}
The occurrence of a gap in the order parameter at temperature
$\tilde{t}<t_c$ is, to our knowledge, a new phenomenon in the theory
of phase transitions.

For $r_{c1}<r<r_{c2}$, the order parameter $m_1\neq 0$ in the
temperature interval $\tilde{t}<t<t_c$. There is no new spontaneous
symmetry breaking with respect to $m_1$ at $t=\tilde{t}$. However,
its jump at $t=\tilde{t}$ is similar to that which occurs in the
usual first-order phase transition scenario. Similar behaviour is
observed at $t=\tilde{t}$ for the latent heat $\Delta E$ and for
$\Delta m_2$. These functions are shown in Figs.~\ref{fig4}c too.

\begin{figure}[t]
\centerline{\includegraphics[angle=0, width=7cm]{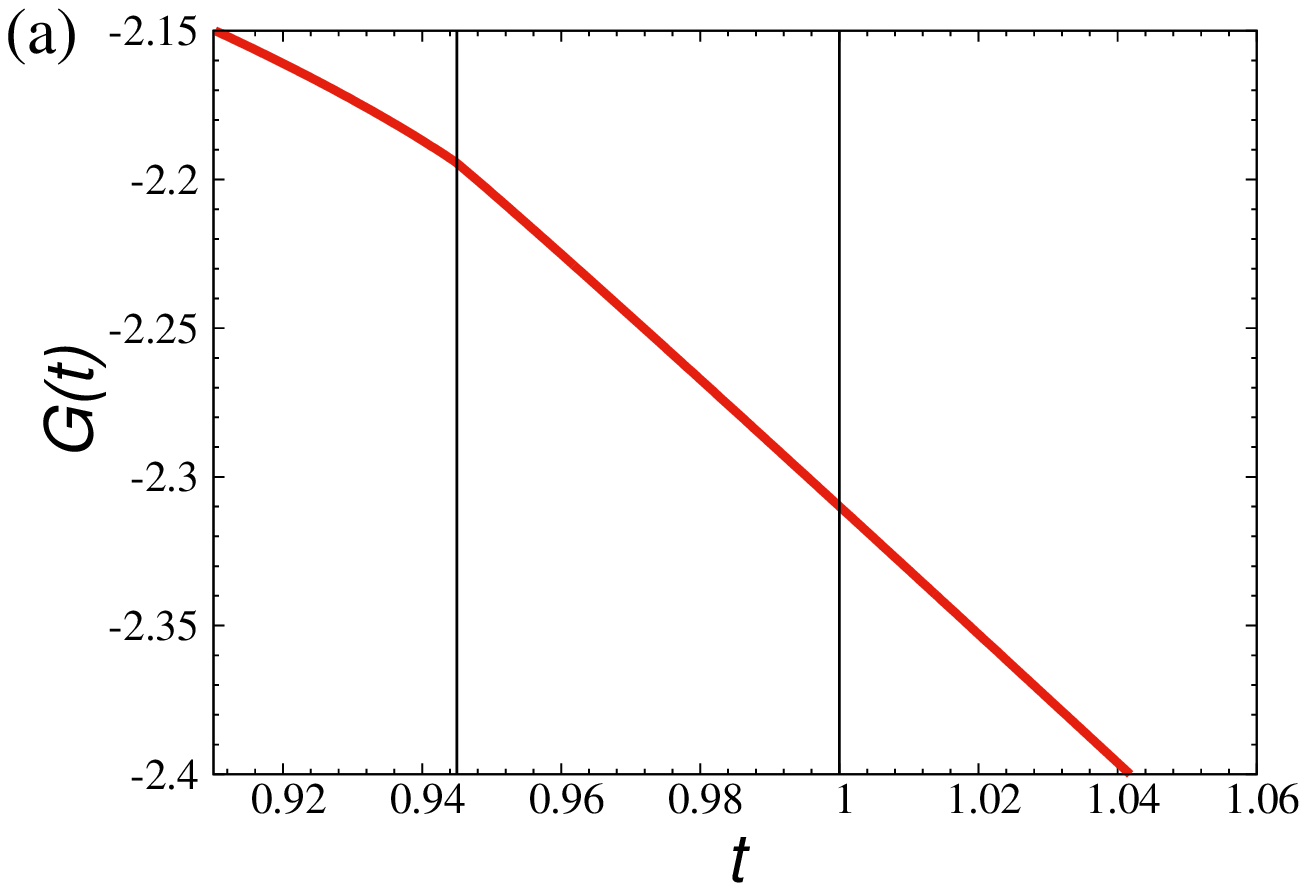}\hspace{1cm}
\includegraphics[angle=0, width=7cm]{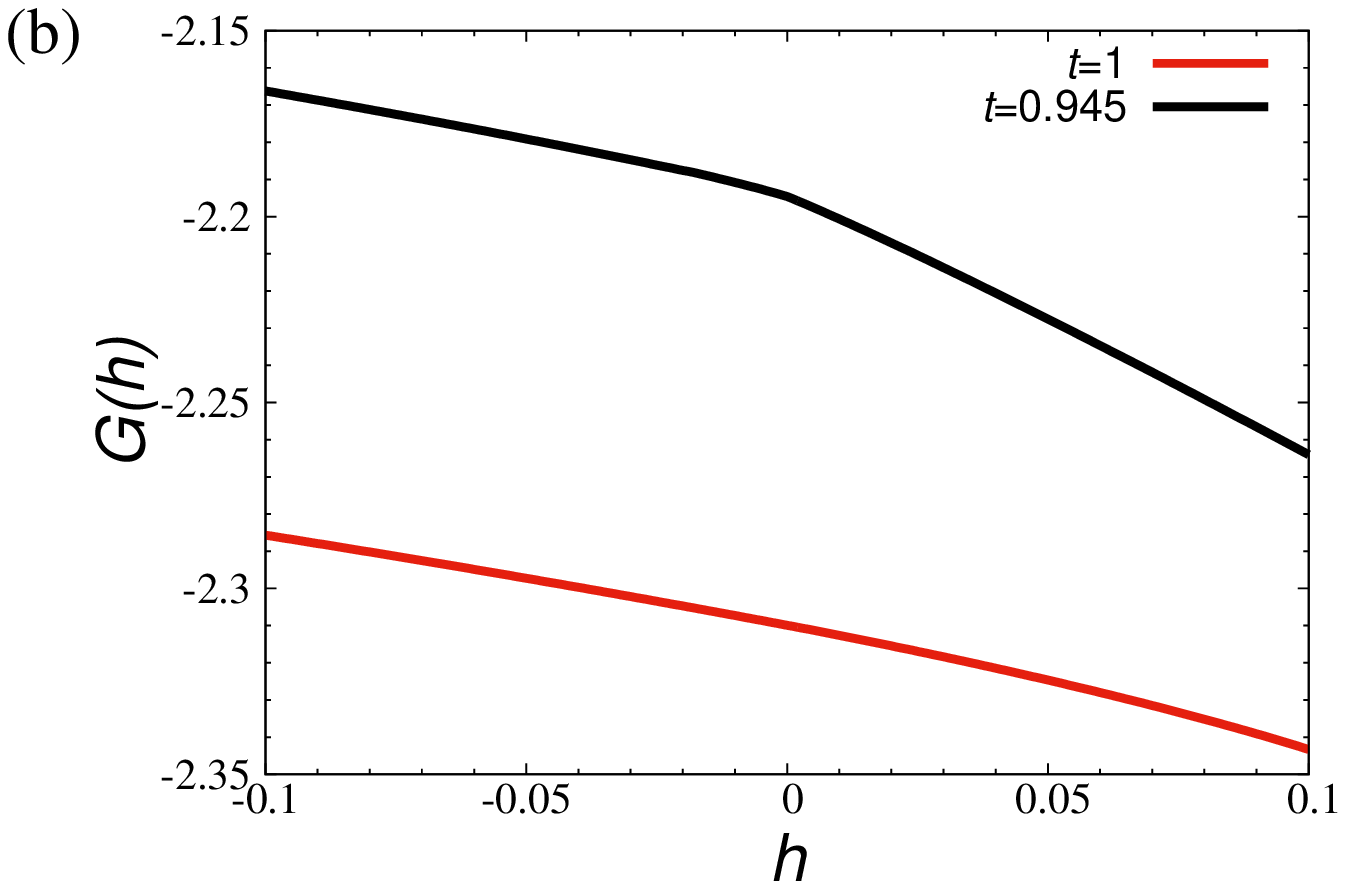}}
\vspace{1ex} \centerline{\includegraphics[angle=0,
width=7cm]{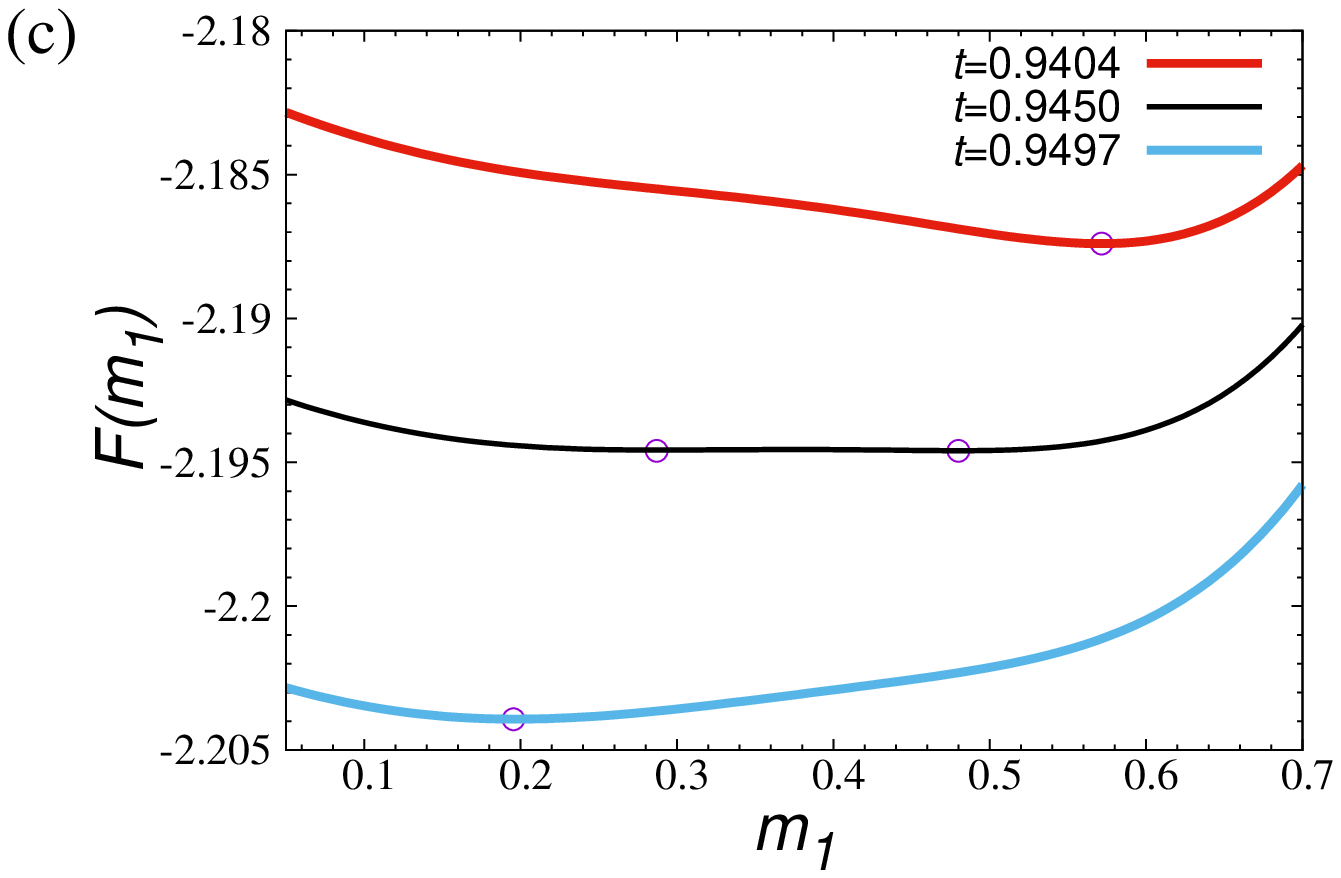}\hspace{1cm}
\includegraphics[angle=0, width=7cm]{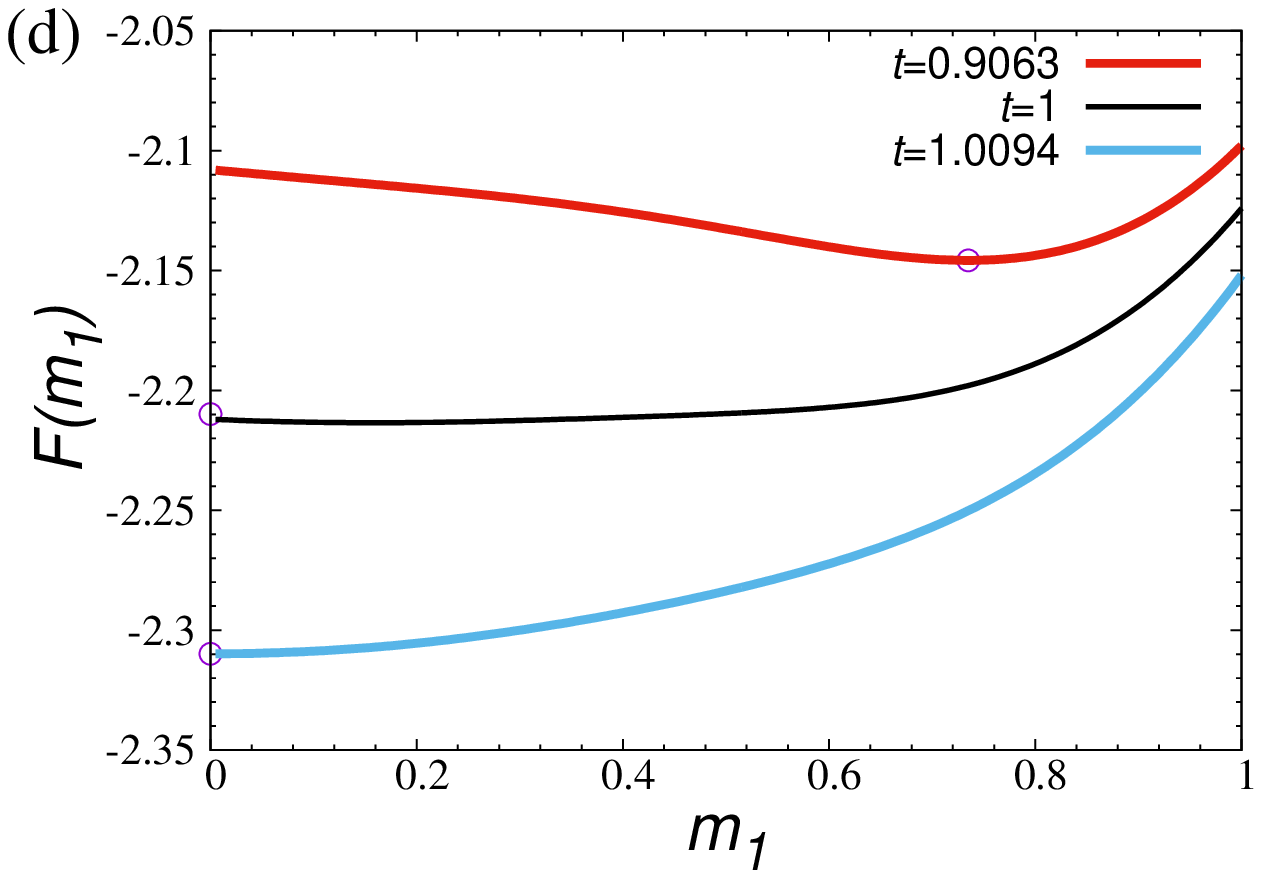}}
\vspace{1ex} \caption{{Typical behaviour of the free energy of the
Potts model with invisible states at $1\leq q <2$ and $r_{c1} < r <
r_{c2}$ ($q=1.2$ and $r=6.96$ for this figure); (a): Gibbs free
energy $G(t)$ at $h=0$. Temperatures $\tilde{t}$ and $t_c$ are shown
by vertical lines. One sees a cusp at  $\tilde{t}$, it signals about
the presence of a latent heat (a jump in the entropy at
$\tilde{t}$); (b): Gibbs free energy $G(h)$ at $t=\tilde{t}$ (upper
curve) and at $t=t_c$ (lower curve). A cusp in the upper curve
signals about a jump in the order parameter at $\tilde{t}$. Note
that the cusp is absent in lower curve: the order parameter is
continuous at $t_c$; (c): Mean-field free energy $F(m_1)$
at $h=0$ and $t<\tilde{t}$, $t\simeq\tilde{t}=0.945$, $t>\tilde{t}$
(upper, middle, and lower curves, correspondingly); (d): Landau free
energy $F(m_1)$ at $h=0$ and $t<t_c$, $t=t_c=1$, $t>t_c$. The
circles in Figs. (c) and (d) show global minima of the free energy.}
 \label{fig5}}
\end{figure}

The behaviour of the Gibbs free energy $G(t,h)$ of the Potts model
with invisible states in the region $1\leq q <2$ and $r_{c1} < r <
r_{c2}$  is elucidated in Figs.~\ref{fig5} a,  b. To this end, we
used conditions (\ref{12a}), (\ref{12b}) to eliminate the
order-parameter dependency of the mean-field free energy (\ref{11})
in favour of the external field. In Fig.~\ref{fig5}a we display the
zero-field Gibbs free energy $G(t,h=0)$ as a function of the reduced
temperature $t$. The cusp at $t=\tilde{t}$ signals  the jump in the
entropy. Fig.~\ref{fig5}b shows the Gibbs free energy $G(h)$ at
$t=\tilde{t}$ (upper curve) and at $t=t_c$ (lower curve). Again, a
cusp in the upper curve signals a jump in the order parameter at
$\tilde{t}$. However, the cusp is absent in the lower curve: the
order parameter is continuous at $t_c$. The mean-field free energy
$F(m_1,m_2)$ is further analysed in Figs.~\ref{fig5}c, d. There we
show the typical behaviour of the free energy as a function of the
first-order parameter $m_1$ at $h=0$ in the region of temperatures
in the vicinity of $t=\tilde{t}$ (c) and $t=t_c$ (d). To get
two-dimensional plots, parameter $m_2$ has been excluded from the
minimum conditions (\ref{12a}), (\ref{12b}). Fig. (c) demonstrates
behaviour typical for the first-order phase transition: two minima
exist at $t=\tilde{t}=0.945$, see the middle curve of the figure.
Different situation is observed in Fig. (d). There, the only value
$m_1=0$ corresponds to the free energy minimum at $t=t_c=1$.

\begin{figure}[t]
    \centerline{\includegraphics[angle=0, width=8cm]{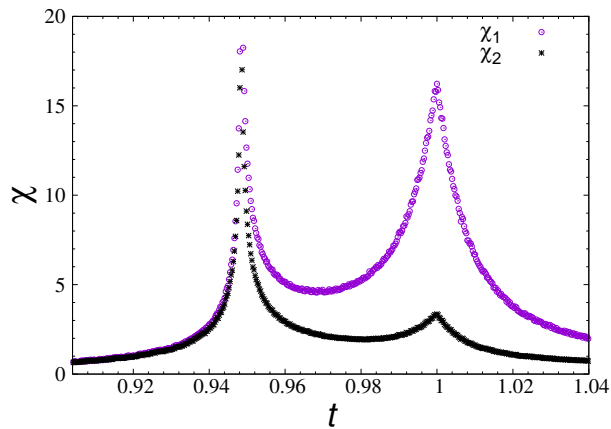}}
        \vspace{1ex}
   \caption{Typical behaviour of the isothermal susceptibilities $\chi_{1}$
 and $\chi_{2}$ as functions of reduced temperature
   for $1< q <2$, $ r_{c1} < r <r_{c2}$ ($q=1.2$ and $r=6.96$ in this figure).
   Distinct peaks are observed at $\tilde{t}$ and $t_c$. \label{fig6}}
\end{figure}

To better understand  temperature behaviour of the order parameters
in the vicinity of $\tilde{t}$, we present, in Fig.~\ref{fig6}
typical plots of the isothermal susceptibilities
${\displaystyle{\chi_{1}={{{\partial m_1}/{\partial h}}}}}$    and
${\displaystyle{ \chi_{2}={{\partial m_2}/{\partial h}}}}$ for $1< q
<2$ and $ r_{c1} < r <r_{c2}$ (specifically, $q=1.2$ and $r=6.96$ in
this figure). One observes two distinct peaks located at $\tilde{t}$
and $t_c$. The values of the susceptibilities were obtained by
numerical evaluation of derivatives in the limit $h\to 0$.

\begin{table}[b]
    \centering \caption{Marginal dimensions $r_{c1}$, $r_{c2}$
        for different values of $1\leq q \leq 2$.
        \label{tab2}}
        \begin{tabular}{|l|l|l|}
        \hline
        $q$   & $r_{c1}$&$r_{c2}$\\ \hline
        1     &7.334(49)    &9.55(35) \\ \hline
        1.1   &7.132(7)&8.995(5)\\ \hline
        1.2   &6.834(11)&8.495(5)\\ \hline
        1.3   &6.577(5)&8.025(5)\\ \hline
        1.4   &6.268(9)&7.535(5)\\ \hline
        1.5   &5.980(6)&7.025(5)\\ \hline
        1.6   &5.658(7)&6.525(5)\\ \hline
        1.7   &5.315(5)&6.025(5)\\ \hline
        1.8   &4.914(8)&5.505(5)\\ \hline
        1.9   &4.447(9)&4.825(5)\\ \hline
        2     &3.622(8)&3.65(5)\\ \hline
    \end{tabular}
\end{table}

\begin{figure}[t]
\centerline{\includegraphics[angle=0, width=8cm]{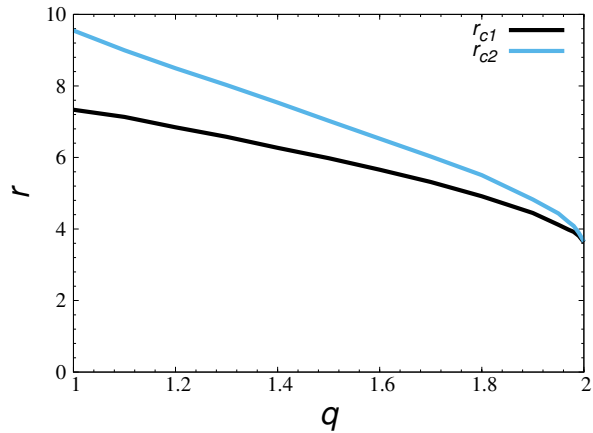} }
\caption{Marginal dimensions $r_{c1}$ (lower curve) and $r_{c_2}$
(upper curve) for the Potts model at $1\leq q \leq 2$. At $q=2$ both
$r_{c1}$ and $r_{c_2}$ coincide within the fidelity interval:
$r_{c1}=r_{c2}\simeq 3.65(5)$. \label{fig7}}
\end{figure}

Values of the marginal dimensions $r_{c1}$ and $r_{c2}$ for
different $q$ are collected in Table~\ref{tab2}. We give the average
value of $r_{c1}$ obtained numerically from the behaviour of the
functions $\Delta m_1$, $\Delta m_2$, and $\Delta E$. The estimate
for $r_{c2}$ has been obtained from the behaviour of $m_1$ as the
minimal value of $r$ for which condition (\ref{cond1}) holds.
Fig.~\ref{fig7} shows the $q$-dependencies of $r_{c1}$ and $r_{c2}$.
For the case $q=2$, where both marginal dimensions $r_{c1}$ and
$r_{c2}$ have to coincide, we use the  estimate $r_c=3.65(5)$ since
it includes both values quoted in the table. It is worth noting,
that in the region $1 \leq q \leq 2$ the difference in the marginal
dimensions is nicely approximated by a linear function:
$r_{c2}-r_{c1}\simeq 2(2-q)$,  although we do not have a simple
explanation for this observation.

In the limiting case $q\to 1$, Eq.~(\ref{11}) gives the free energy
that depends on the second-order  parameter $m_2$ only:
\begin{equation}
\lim_{q\to 1} f(m_1,m_2)
 = \frac{z \left(m_2
r+1\right)^2}{2(r+1)^2}-T \log \left(e^{\frac{h+\frac{z  (m_2 r
+1)}{r+1}}{T}}+r\right).
\end{equation}
Minimizing the free energy with respect to $m_2$ one gets the
temperature behaviour $m_2(T)$. In turn, the appearance of a gap in
this dependence can be used as a condition to determine the marginal
dimension $r=r_{c1}$. We estimate numerically the marginal
dimension $r_{c2}$ from the limit $\lim_{q\to 1^+}r_{c2}(q)$.

\section{Conclusions}

In this paper, we reconsidered the mean field approach to the
recently introduced Potts model with invisible states
\cite{Tamura10}. The model was suggested in an attempt to resolve
controversies between theoretical predictions of a second-order
phase transition and simulations indicating  a first-order scenario
\cite{example}. Indeed, the number  of invisible states $r$
introduced into the model plays the role of a parameter that
regulates the order of the phase transition. In the Potts case with
$q$ interacting (visible) states, the transition becomes the first
order starting from a certain, marginal, value $r_c$. For the
particular case $q=2$ it had already been shown that the change
occurs at large $r$ (for $d=2$) and that $3< r_c<4$ within the mean
field analysis \cite{Tamura10,Tamura12}. Here, we determine a more
precise estimate for the value for $q=2$, namely $r_c\simeq
3.65(5)$. This estimate is in excellent agreement with the $z\to
\infty$ limit of the result obtained for the Bethe lattice  with $z$
nearest neighbours \cite{Ananikian13}.

We then considered the region $1 \leq q<2$. There, the mechanism
which changes the order of the phase transition is, to our
knowledge,  new to the field. Namely, there are two marginal
dimensions,  $r_{c1}$ and $r_{c2}$, as indicated in
Fig.~\ref{fig4}a,  b. They characterise the temperature behaviour of
the first derivatives of the free energy. At $r_{c1}$ the  gap
appears at a certain temperature $\tilde{t}<t_c$, however, the
transition at $t_c$ remains of  second order. The gap increases and
moves towards $t_c$ with further increase of $r$, and, finally at
$r_{c2}$ it reaches $t_c$: the transition becomes the first order.
The marginal dimensions $r_{c1}$ and $r_{c2}$ are functions of $q$,
they are shown in Fig.~\ref{fig7} and listed in Table~\ref{tab2}.

The above observations shows essential difference in behaviour of
the Potts model with invisible states at $q=2$  and at $1\leq q <2$.
Whereas in the former case an increase of $r$ turns the second-order
phase transition to the weak first-order transition (the jump in the
order parameter $\lim_{r\to r_{c}} \Delta m_1 \to 0$), in the latter
case the second transition is turned to the sharp first-order
transition, $\lim_{r\to r_{c2}} \Delta m_1 \neq 0$. The same
conclusion can be reached analysing the latent heat behaviour. In
this respect, it is worth recalling that the case $q=1$, considered
here, corresponds to the bond percolation. In turn, our result means
that adding invisible states turns transition in the
bond-percolation model into strong first-order. It is also worth
mentioning here other mechanisms that are known to sharpen
percolation transitions, such as those delivering explosive
\cite{Achlioptas09,Bastas14} or bootstrap \cite{Adler91}
percolation.

Marginal dimensions are widely met when phase transitions are
analysed. Some examples are mentioned in the beginning of this
paper. The first step  towards defining these quantities for the
Potts model with invisible states have been taken. More involved
theories  have to take into account fluctuations, which are wiped
out within the method we currently used. One such approach, the
renormalization group, leads to a perturbative expansion where the
mean-field result enters as a first term. It is instructive to note
here, that already this first-order contribution is non-integer, as
foreseen by the results of this paper. Moreover, usually one expects
logarithmic corrections to scaling to accompany the change in the
order of a phase transition \cite{Kenna13}. Again, these topics are
out of reach of the mean-field analysis and will be a subject of a
separate study.

\vspace{1cm}

This work was supported in part by FP7 EU IRSES projects  No.
$295302$ ``Statistical Physics in Diverse Realizations", No.
$612707$ ``Dynamics of and in Complex Systems", No. $612669$
``Structure and Evolution of Complex Systems with Applications in
Physics and Life Sciences", and by the Doctoral College for the
Statistical Physics of Complex Systems,
Leipzig-Lorraine-Lviv-Coventry $({\mathbb L}^4)$.

\section*{References}

\providecommand{\newblock}{}


\begin{thebibliography}{10}
\expandafter\ifx\csname url\endcsname\relax
  \def\url#1{{\tt #1}}\fi
\expandafter\ifx\csname urlprefix\endcsname\relax\def\urlprefix{URL
}\fi \providecommand{\eprint}[2][]{\url{#2}}

\bibitem{Stanley88}
Stanley H E (1988) {\em Introduction to Phase Transitions and
Critical Phenomena (International Series of Monographs on Physics)}
(Oxford Science Publications, Oxford)

\bibitem{Domb96}
Domb C, {\em The Critical Point} (Taylor \& Francis, London, 1996)


\bibitem{Stanley69}
Stanley H E 1969 {\em Phys. Rev.} {\bf 179} 570

\bibitem{Mermin66}
Mermin  N D and Wagner H  1966 {\em Phys. Rev. Lett.} {\bf 17} 1133

\bibitem{Hohenberg67}
Hohenberg P C  1967 {\em Phys. Rev.} {\bf 158}  383

\bibitem{Aharony76}
Aharony A 1976, in {\em Phase Transitions and Critical Phenomena},
edited by Domb C and Green M S, Academic Press, London {\bf 6}

\bibitem{cubic}
Sznajd J 1984 {\em J. Magn. Magn. Mater.} {\bf 42} 269; Doma\'{n}ski
Z and Sznajd J 1985 {\em Phys. Status Solidi B}  {\bf 129} 135

\bibitem{Folk00}
Kleinert H and Schulte-Frohlinde V  1995 {\em Phys. Lett. B} {\bf
342} 284; Folk R, Holovatch Yu, and Yavors'kii T 2000 {\em Phys.
Rev. B} {\bf 62} 12195,     {\em Erratum} {\em Phys. Rev. B} {\bf
63} 189901(E)

\bibitem{Dudka04}
Dudka M, Holovatch Yu, and Yavors'kii T 2004 {\em J. Phys. A: Math.
Gen.} {\bf 37} 10727

\bibitem{Potts52}
Potts R B 1952 {\em Proc. Camb. Phil. Soc.} {\bf 48} 106

\bibitem{Wu82}
Wu F Y 1982 {\em Rev. Mod. Phys.} {\bf 54} 23

\bibitem{non-colinear}
Holovatch Yu, Ivaneyko D, and Delamotte B 2004 {\em J. Phys. A:
Math. Gen.} {\bf 37} 3569

\bibitem{frustrations}
Delamotte B, Mouhanna D, and Tissier M 2004 {\em Phys. Rev. B}  {\bf
69} 134413; Calabrese P and Parruccini P  2004 {\em Nucl. Phys. B}
{\bf 679} 568

\bibitem{structural_disorder}
Pelissetto A and Vicari E 2002 {\em Phys. Rep.} {\bf 368} 549;
 Folk R, Holovatch Yu,  and Yavors'kii T 2003 {\em Physics-Uspiekhi} {\bf 46} 169

\bibitem{competing_fields}
Halperin B I, Lubensky T C, and Ma S 1974 {\em Phys. Rev. Lett.}
{\bf 32} 292; Folk R and Holovatch Yu 1996 {\em J. Phys. A: Math.
Gen} {\bf 29} 3409; Dudka M, Folk R, Holovatch Yu, and Moser G 2012
{\em Condens. Matter Phys.} {\bf 15} 43001



\bibitem{Tamura10}
Tamura R, Tanaka S, and Kawashima N 2010 {\em Prog. Theor. Phys.}
{\bf 124} 381

\bibitem{example}
Tamura R and Kawashima N 2008 {\em J. Phys. Soc. Jpn.} {\bf 77}
103002; Stoudenmire E M, Trebst S, and Balents L 2009 {\em Phys.
Rev. B} {\bf 79} 214436; Okumura S, Kawamura H, Okubo T, and Motome
Y 2010 {\em J. Phys. Soc. Jpn.} {\bf 79} 114705

\bibitem{Tanaka11b}
Tanaka S, Tamura R, and Kawashima N 2011 {\em J. Phys.: Conf. Ser.}
{\bf 297}  012022


\bibitem{Tamura12}
Tamura R, Tanaka S, and Kawashima N  2011 {\em arXiv:1111.6509}

\bibitem{Johnston13}
Johnston D A and  Ranasinghe R P K C M 2013 {\em J. Phys. A: Math.
Theor.} {\bf 46}  225001

\bibitem{Tanaka11a}
Tanaka S and Tamura R 2011 {\em J. Phys.: Conf. Ser.} {\bf 320}
012025

\bibitem{Mori12}
Mori T 2012 {\em J. Stat. Phys.} {\bf 147}  1020

\bibitem{Enter11a}
van Enter A C D, Iacobelli G, and Taati S 2011 {\em Prog. Theor.
Phys.} {\bf 126}  983

\bibitem{Enter11b}
van Enter A C D, Iacobelli G, and Taati S 2011 {\em
arXiv:1109.0189}

\bibitem{Ananikian13}
Ananikian N, Izmailyan N Sh, Johnston D A, Kenna R, and  Ranasinghe
R P K C M 2013 {\em J. Phys. A: Math. Theor.} {\bf 46}  385002

\bibitem{mfa}
Kihiara T, Midzuno Y, and Shizume J 1954 {\em J. Phys. Soc. Jpn.}
{\bf 9} 681; Mittag L and Stephen J 1974 {\em J. Phys. A: Math.
Gen.} {\bf 7} L109; Straley J P and Fisher M E 1973 {\em J. Phys. A:
Math. Gen.} {\bf 6} 1310

\bibitem{Fortiun69}
Kasteleyn P W and Fortuin C M 1969 {\em J. Phys. Soc. Jpn.}  {\bf
26} (Suppl.) 11

\bibitem{Giri77}
Giri M R, Stephen M J, and Grest G S 1977 {\em Phys. Rev. B} {\bf
16} 4971


\bibitem{Stephen76}
Stephen M J  1976 {\em Phys. Lett. A} {\bf 56} 149

\bibitem{Deng}
Deng Y, Garoni T M, and Sokal A D 2007 {\em Phys. Rev. Lett.}  {\bf
98} 030602

\bibitem{Jacobsen05}
Jacobsen J L and Saleur H 2005 {\em Nucl. Phys. B} {\bf 716} 439

\bibitem{Fortiun72}
Fortuin C M and Kasteleyn P W 1972 {\em Physica}  {\bf 57} 536

\bibitem{Stenull}
Stenull O, Janssen H K, and Oerding K 1999  {\em Phys. Rev. E} {\bf
59} 4919

\bibitem{Aharony78}
Aharony A 1978 {\em J. Phys. C: Solid State Phys.} {\bf 11}

\bibitem{Aharony79}
Aharony A and Pfeuty P 1979 {\em J. Phys. C: Solid State Phys.}
{\bf 12}

\bibitem{Siva}
Siva K, Tao J, and Marcolli M 2015 {\em arXiv: 1508.00504}

\bibitem{Lubensky78}
Lubensky T C and Isaacson J  1978{\em Phys. Rev. Lett.} {\bf 41} 12

\bibitem{Achlioptas09} Achlioptas D,  D'Souza R M, and
Spencer J 2009 {\em Science} {\bf 323} 5920

\bibitem{Bastas14} Bastas N, Giazitzidis P, Maragakis M, and Kosmidis K
2014 {\em Physica A} {\bf 407} 54

\bibitem{Adler91}
Adler J 1991 {\em Physica A} {\bf 171} 453

\bibitem{Kenna13}
Kenna R 2013 in: {\em Order,  Disorder  and  Criticality} vol 3),
ed. Holovatch Yu (Singapore:  World Scientific) pp 1-46


\end{thebibliography}
\end{document}